\documentclass[conference,letterpaper]{IEEEtran}
\IEEEoverridecommandlockouts
\usepackage{cite}
\usepackage[nolist]{acronym}
\usepackage{amsmath,amssymb,amsfonts}
\usepackage{algorithmic}
\usepackage{graphicx}
\usepackage{textcomp}
\usepackage{xcolor}
\usepackage{hyperref}
\usepackage{caption}
\usepackage{enumitem}
\usepackage{lipsum}
\usepackage{colortbl}
\usepackage{epstopdf}
\usepackage{listings}

\usepackage{todonotes}
\usepackage{pdfsync}

\newcolumntype{G}{>{\columncolor[gray]{0.8}[.5\tabcolsep][\tabcolsep]}c}

\newcommand{\muops}{\mbox{$\mu$-ops}}

\begin{acronym}
    \acro{ecm}[ECM]{Execution-Cache-Memory}
    \acro{mca}[llvm-mca]{LLVM Machine Code Analyzer}
    \acro{xgss}[llvm-exegesis]{LLVM Machine Instruction Benchmark}
    \acro{cqa}[CQA]{Code Quality Analyzer}
    \acro{ufs}[UFS]{Uop Flow Simulation}
\end{acronym}

\def\BibTeX{{\rm B\kern-.05em{\sc i\kern-.025em b}\kern-.08em
    T\kern-.1667em\lower.7ex\hbox{E}\kern-.125emX}}
\begin{document}

\title{{Automatic Throughput and Critical Path Analysis of x86 and ARM Assembly Kernels}
\thanks{This work was in part funded by the BMBF project METACCA.}}

\author{\IEEEauthorblockN{
Jan Laukemann, Julian Hammer, Georg Hager and Gerhard Wellein}
\IEEEauthorblockA{
\{jan.laukemann, julian.hammer, georg.hager, gerhard.wellein\}@fau.de\\
Erlangen Regional Computing Center\\
Friedrich-Alexander-Universit\"at Erlangen-N\"urnberg, Erlangen, Germany}}

\maketitle

\begin{abstract}
Useful models of loop kernel runtimes on out-of-order architectures require an analysis of the in-core performance behavior of instructions and their dependencies.
While an instruction throughput prediction sets a lower bound to the kernel runtime, the critical path defines an upper bound.
Such predictions are an essential part of analytic (i.e., white-box) performance models like the Roof{}line and \ac{ecm} models. They enable  a better understanding of the performance-relevant interactions between hardware architecture and loop code.

The Open Source Architecture Code Analyzer (OSACA) is a static analysis tool for predicting the execution time of sequential loops. It previously supported only x86 (Intel and AMD) architectures and simple, optimistic full-throughput execution. We have heavily extended OSACA to support ARM instructions and critical path prediction including the detection of loop-carried dependencies, which turns it into a versatile cross-architecture modeling tool. We show runtime predictions for code on Intel Cascade Lake, AMD Zen, and Marvell ThunderX2 micro-architectures based on machine models from available documentation and semi-automatic benchmarking. The predictions are compared with actual measurements.
\end{abstract}


\section{Introduction}\label{sec:intro}

Analytic performance modeling of compute-intensive applications during development or optimization can be a powerful tool and sheds light on how code executes on modern CPU architectures. Such models are hence not only constructed for the sake of prediction but also to study relevant bottlenecks and to assess the compiler's ability to generate optimal code.
However, they require a deep understanding of the underlying micro-architecture in order to yield accurate results.
Common (simplified) approaches for numerical kernels are the Roof{}line~\cite{Roofline} model or the \ac{ecm}~\cite{ECM} model, whose construction is supported by the \emph{Kerncraft} open-source performance modeling tool~\cite{Kerncraft}. For Roof{}line, the Roof{}line Model Toolkit~\cite{RooflineModelToolkit} and Intel's Roof{}line Advisor\footnote{\url{https://software.intel.com/en-us/advisor-user-guide-roofline-analysis}} are also available.

In general, there are two approaches to predict runtime and performance behavior: simulation and static analysis. Our work implements the latter.
Even though simulators may be more thorough and accurate if comprehensive implementations exist, their usage is complicated by obstacles like finding steady states for throughput analysis and pinpointing the runtime-defining hardware bottleneck. In addition, their implementation is much more complex.
The analysis and modeling process is split into in-core execution time and data transfer time through the memory hierarchy. See~\cite{Kerncraft, ECM} for examples on how this is done.
For a long time, the only capable tool for static in-core code analysis was Intel's Architecture Code Analyzer (IACA)~\cite{iaca}, which was employed by Kerncraft.
Besides being at end-of-life, there are multiple limitations: Intel-only architectures, undisclosed model and later versions restricted to full-throughput assumption.
To improve on this, we develop the \emph{Open Source Architecture Code Analyzer} (OSACA)~\cite{Osaca}, which has, in addition to the features already known from IACA, extended x86 (Intel Cascade Lake and AMD) and AArch64 ARM support and supports critical path (CP) latency analysis and loop-carried dependency detection.
All three predictions can be combined to a more accurate performance model, including the throughput as a lower bound and the critical path as an upper bound of the kernel runtime. Like IACA, OSACA assumes that all data originates from the first-level cache (i.e., L1 cache).

With OSACA's semi-automatic benchmarking pipeline, compilers can benefit from an automated model construction~\cite{RooflineModelToolkit,Kerncraft}. The instruction database is dynamically extendable, which enables users to adapt the tool to other application scenarios beyond numerical kernels found in HPC usecases.

This paper is organized as follows: In Section~\ref{ssec:related}, we cover related work.
Section~\ref{sec:methodology} details the model assumptions and construction for the underlying architecture and the general methodology of the throughput and critical path analysis as well as the loop-carried dependency detection.
In Section~\ref{sec:results} we describe the benchmarking hardware/software environment and validate the methodology against actual measurements and compare with related tools.
Section~\ref{sec:conclusion} summarizes the work and gives an outlook to future developments.

The OSACA software is available for download under AGPLv3~\cite{osaca-git}. Information about how to reproduce the results in this paper can be found in the artifact description~\cite{artif}.

\subsection{Related Work}\label{ssec:related}
OSACA was inspired by IACA, the Intel Architecture Code Analyzer~\cite{iaca}.
Developed by Israel Hirsh and Gideon S. [sic], Intel released the tool in 2012 and announced its end-of-life in April 2019.
Therefore, no feature enhancements or new microarchitecture support can be expected.
It is closed source and the underlying model has neither been published by the authors, nor peer reviewed.
The latest version supports throughput analysis on Intel micro-architectures up to Skylake (including AVX-512), but is not capable of critical path analysis or loop-carried dependency detection.

\ac{mca}~\cite{llvm_mca} is a performance analysis tool based on LLVM's existing scheduling models.
Currently it lacks support for HPC-relevant ARM architectures such as the ThunderX2, and some scheduling models need refinement. Also, \ac{mca} cannot analyze CPs, even though one can manually identify a CP by the provided latency analysis.
\ac{xgss}~\cite{llvm_exegesis} is a micro-benchmarking framework for measuring throughput and latency of instruction forms. It could thus be used as a data source to feed the OSACA instruction database.
Mendis et al.~\cite{ithemal-icml} apply a deep neural network approach to estimate block throughput on Intel x86 architectures from Ivy Bridge to Skylake.
It is able to use IACA byte markers for indicating the code block to analyze and is currently not capable of detecting CPs or loop-carried dependencies.
\ac{cqa}~\cite{cqa} is a static performance analysis tool focused on single-core performance of loop-centric x86 code.
Unlike OSACA, its goal is not to predict runtime, but rather give the developer a quality estimate of the code based on static binary analysis.
\ac{ufs}~\cite{ufs} extends \ac{cqa} with a simulator for the out-of-order execution, modeling aspects OSACA assumes to be based on fixed (non-optimal) probabilities.

There are a fair number of simulators available: gem5, developed by Binkert et al.~\cite{gem5}, ZSim by Sanches et al.~\cite{zsim} and MARSSx86 by Patel et al.~\cite{MARSSx86}.
While gem5 even supports various non-x86 instruction set architectures (ARM, Power, RISC-V among others), all of them are considered as ``full-system'' simulators, going above and beyond the scope of this work.
Therefore, they give a coarse overview on complete (multi- or many-core) systems, rather than detailed insights pinpointing a bottleneck.

\section{Methodology}\label{sec:methodology}

When analyzing loop kernels, we assume for each CPU architecture a corresponding ``port model'':
Each assembly instruction is (optionally) split into \emph{micro-ops} (\muops), which get executed by multiple ports.
A particular instruction may have multiple ports that can execute it (e.g., two integer ALUs), or -- in case of complex instructions -- multiple ports that \emph{must} execute it (e.g., combined load and floating-point addition). Shared resources, such as a divider pipeline or a data load unit, are modelled as additional ports.

Each port receives at most one instruction per cycle and may be blocked by an instruction for any number of cycles. To model parallel execution of the same instruction form on multiple ports, the cycles may be spread among multiple ports, also allowing the inverse of integers as acceptable cycle throughput of an instruction per port, but always adding up to at least one cycle per instruction over all ports.

Both x86 and ARM allow memory references to be used in combination with arithmetic instructions. This is modelled by 
splitting the instruction in the load and the arithmetic part, and accounting for their respective port pressures and dependencies separately (see below). 

Figure~\ref{fig:generic_portmodel} shows a diagram of the generic port model. Cascade Lake would be modeled with eight ports, plus one divider pipeline port and two data ports. A floating-point divide instruction would occupy port 0 for one cycle and the DIV port (i.e., pipeline) for four cycles, while an add instruction would use ports 0 and 1 for each half a cycle, because it may be executed on both.

\begin{figure}[tbp]
\centerline{\includegraphics[width=0.45\columnwidth]{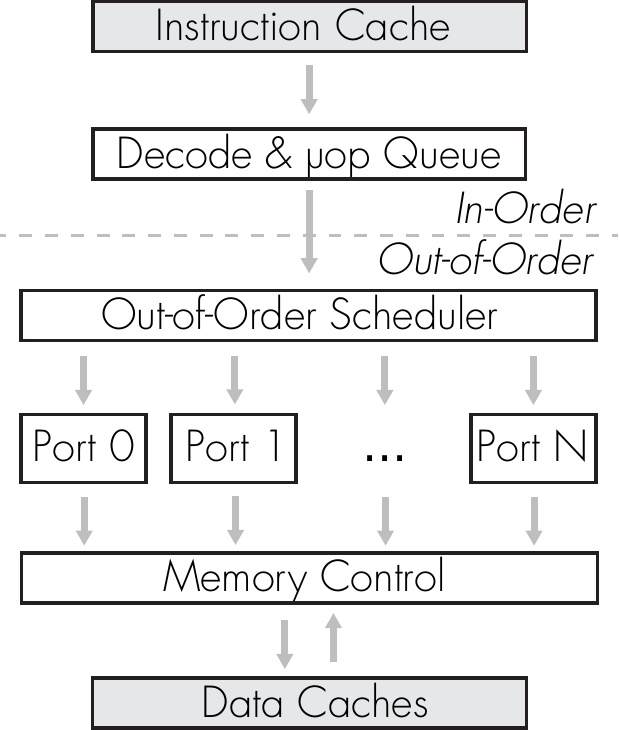}}
\caption{Assumed generic out-of-order port model. Other shared resources (e.g., DIV pipeline) are modeled as additional ports.}
\label{fig:generic_portmodel}
\end{figure}

We repeat here the assumptions behind our prediction model~\cite{Osaca}:
\newline
\begin{itemize}
    \item \emph{All data accesses hit the first-level cache.}
          This is where the boundary between in-core and data transfer analysis is drawn.
          If a dataset fits in the first-level cache, no cache misses occur. Replacement strategies, prefetching, line buffering, etc., are insignificant on this level.
          Behavior beyond L1 can be modeled with Kerncraft~\cite{Kerncraft}, which relies on an in-core analysis from OSACA and combines it with data analysis to arrive at a unified model prediction.
    \item \emph{Multiple available ports per instruction are utilized with fixed probabilities.}
          If the exact amount of \muops\ per port per instruction form is unknown, we assume that all suitable ports for the same instruction are used with fixed probabilities. E.g., an \texttt{add} instruction that may
          use one out of four possible ports and has a maximum throughput of 1\,instr./cy on any unit will be assigned 0.25\,cy on each of the four ports.
          This implies imperfect scheduling if ports are asymmetric.
          Asymmetry means that multiple ports can handle the same instruction, but other features of those ports differ (e.g., one port supports \texttt{add} and \texttt{div}, while another supports \texttt{add} and \texttt{mul}).
          This may cause load imbalance since, e.g., a code with only \texttt{add} and \texttt{mul} may be imperfectly scheduled.
          The consideration of the full kernel for a more realistic port pressure model is currently not supported, but is taken into account for future versions.
\end{itemize}
    
\subsection{Port Model Construction}\label{ssec:model_construction}
\begin{figure}[tbp]
\centering
\includegraphics[width=\columnwidth]{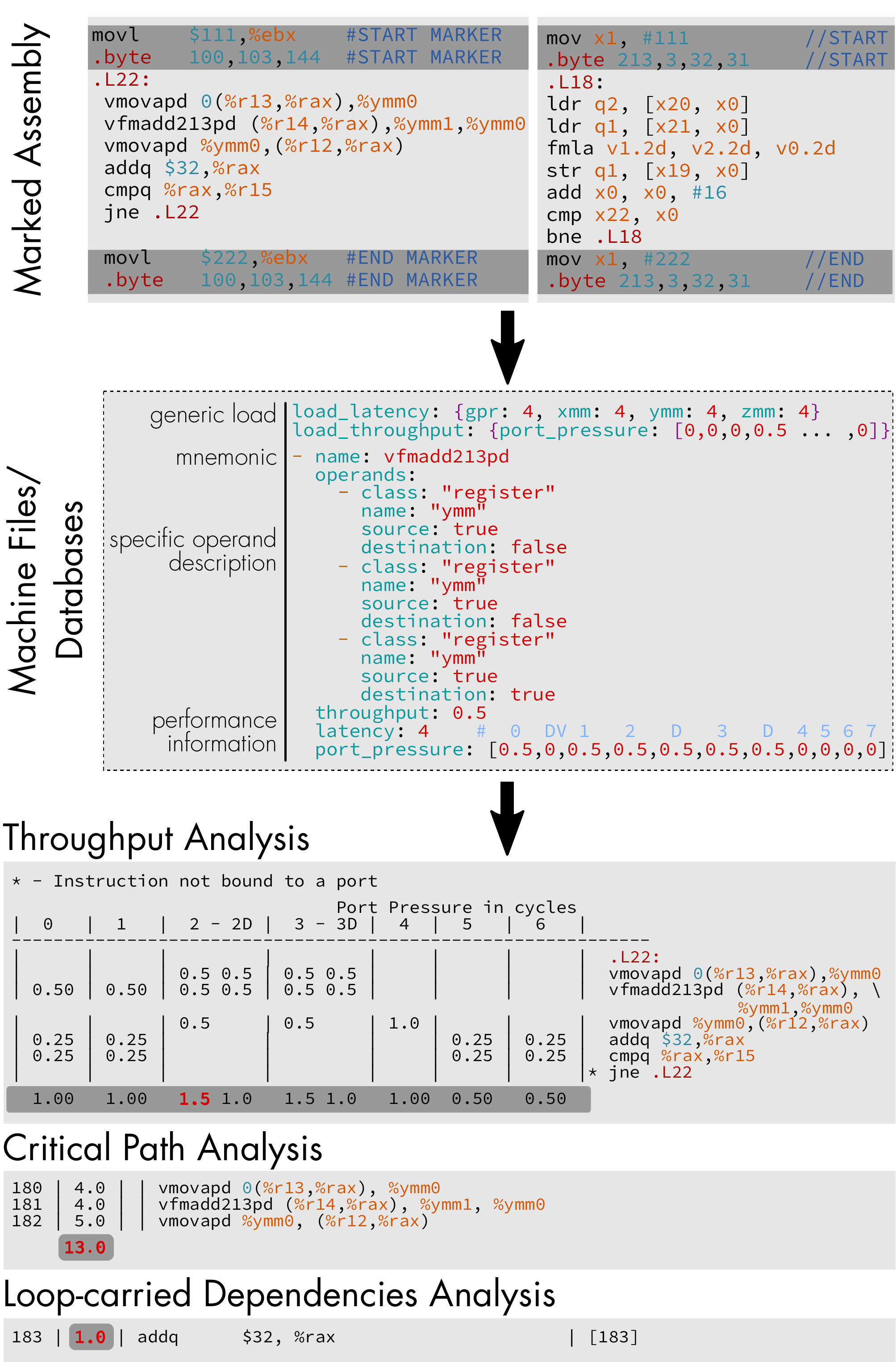}
\caption{Structural design of OSACA and its workflow, for STREAM triad (\texttt{A(:)=B(:)+s*C(:)}) loop.}
\label{fig:methodology}
\end{figure}

The overall methodology of OSACA is exemplified using the STREAM triad \texttt{A(:)=B(:)+s*C(:)} loop in Figure~\ref{fig:methodology}.
The x86 or AArch64 ARM assembly is parsed and the kernel in between the byte markers is extracted.
For convenience, OSACA supports IACA's byte markers for x86 and uses the same instruction pattern for ARM assembly. 
For each parsed instruction form within the kernel, OSACA obtains the maximum inverse throughput and latency in cycles and the ports it can be scheduled to from its instruction database.
Furthermore, it keeps track of source and destination operands for identifying register dependencies.

Possible sources for OSACA's database are microbenchmark databases like uops.info~\cite{Abel19}, Agner Fog's ``Instruction Table''~\cite{Agner}, or specific microbenchmarks using our own frameworks asmbench~\cite{asmjit_poster} and ibench~\cite{ibench}.
For the latter, OSACA can automatically create benchmark files and import the output into its database, resulting in a semi-automatic benchmark pipeline.

\subsection{Instruction Throughput and Latency Analysis} \label{ssec:tplt_analysis}
To obtain the latency and throughput of an instruction, we automatically create assembly benchmarks for use with ibench.
It offers the infrastructure to initialize, run and accurately measure the desired parameters.
It is also intended to support a python-based approach to micro-benchmarking, using the asmbench framework, which is not yet implemented at the time of writing.

Synthetic dependency chain generation within the assembly kernel allows measurement of throughput and latency of an instruction form and has been described in our previous work~\cite{Osaca}.
As stated in Section~\ref{ssec:model_construction}, in addition to directly measuring throughput and latency of instruction forms including memory references in combination with register operands, which currently requires manual effort, OSACA is able to dynamically calculate the throughput by taking the maximum of both the load and arithmetic part and the latency by taking the sum of both parts.
The throughput prediction assumes a fixed and balanced utilization of all suitable ports for any instruction form and perfect out-of-order scheduling without loop-carried dependencies.
It thus yields a lower bound for execution time.

\subsection{Critical Path Analysis}\label{ssec:cp_analysis}
The critical path analysis is based on a directed acyclic graph (DAG) constructed from inter-instruction register dependencies following these rules:
\begin{enumerate}
    \item A vertex is created for every instruction form in the marked piece of code.
    \item From each instruction form's destination operands, edges are drawn to all instruction forms ``further down'' relying on these outputs, unless a break of dependency is found in between (e.g., by zeroing the register).
    \item All edges are weighted with their source instruction's latency.
    \item If a source memory reference has a dependency, an intermediate load-vertex is added along this edge and the additional edge weighted with the load latency.
\end{enumerate}
After creating the DAG, the longest path within it is determined by using a weighted topological sort based on the approach of Manber~\cite{Manber1989IntroductionTA}.
The CP is thus an upper bound for the execution time of a single instance of the loop body.

\subsection{Loop-Carried Dependency Detection}\label{ssec:loop-carried_deps}
Dependencies in between iterations, i.e., loop-carried dependencies (LCDs), can drastically influence the runtime prediction of loop kernels: Even with sufficient out-of-order execution resources, overlap of successive iterations is only possible up to the limit set by the LCD. The actual runtime is thus limited from below by the length of the LCD chain. 
OSACA can detect LCDs by creating a DAG of a code comprising two back-to-back copies of the loop body. It can thus analyze paths from each vertex of the first kernel section and detect most cyclic LCDs if there exists a dependency chain observable by register dependencies from one instruction form to its corresponding duplicate in the next iteration.

\section{Results}\label{sec:results}

\begin{table*}[t]
\centering
\begin{tabular}{l c|r r|r r r|r r r|r r r}
\hline
\rowcolor{white} Architecture & Unroll & \multicolumn{2}{c|}{Measured} & \multicolumn{9}{c}{Prediction [cy/it]} \\
\rowcolor{white}              & factor &        &                      & \multicolumn{3}{c}{OSACA} & \multicolumn{3}{c}{IACA} & \multicolumn{3}{c}{LLVM-MCA} \\
\rowcolor{white}              &        & MLUP/s & cy/it                & TP & LCD & CP         & TP & LCD & CP        & TP & LCD & CP \\
\hline
\hline
Marvel ThunderX2     & 4x & 118.9 & 18.50 & 2.46 & 18.00 & 25.00 & --- & --- & --- & --- & --- & --- \\
Intel Cascade Lake X & 4x & 178.3 & 14.02 & 2.19 & 14.00 & 18.00 & 14.00 & --- & --- & 2.00 & 14.75 & 19.00 \\
AMD Zen              & 4x & 194.4 & 11.83 & 2.00 & 11.50 & 15.00 & --- & --- & --- & 3.00 & 18.00 & 24.00 \\
\hline
\end{tabular}
\caption{Analysis and measurement of the Gauss-Seidel code on three architectures with OSACA, IACA, and LLVM-MCA predictions. Dashes denote unsupported analysis types or architectures. TP is the throughput prediction, a lower runtime bound. LCD is the loop carried dependency prediction, an expected runtime. CP is the critical path prediction, an upper runtime bound.}
\label{tab:meas}
\end{table*}
The CP and LCD detection described in Section~\ref{sec:methodology} are included in OSACA's analysis of loop code and presented together with the ``classic'' throughput results. 
For validation we will use assembly representations generated by the Intel Fortran Compiler for x86 and the GNU Fortran Compiler for ARM, respectively.
In case of CLX we also compare to the IACA and LLVM-MCA predictions for Skylake-X, which does not differ in terms of the port model.
Due to the proprietary nature of IACA, we cannot use it on any AMD- or ARM-based system; hence, we compare against LLVM-MCA on AMD Zen.
For lack of other tools, on TX2 OSACA's prediction can only be compared to measurements.

\subsection{Example: Gauss-Seidel method on CSX, ZEN and TX2}\label{ssec:implementation_example}

An interesting floating-point benchmark for comparing predictions with the measured runtime is a 2D version of the ``Gauss-Seidel'' sweep~\cite{templates}:
\begin{lstlisting}[language={Fortran}, keywordstyle=\color{blue!70!black}\bfseries, commentstyle=\color{blue}, frame=lines, escapechar={ß}]
do it=1,itmax
  do k=1,kmax-1
    do i=1,imax-1
      phi(i,k,t0) = 0.25 * (
        phi(i,k-1,t0) + phi(i+1,k,t0) + 
        phi(i,k+1,t0) + phi(i-1,k,t0))
    enddo                                                  
  enddo
enddo
\end{lstlisting}
It has one multiplication and three additions per iteration.
As the update of the matrix happens in-place, each iteration is dependent on the previously calculated value of its ``left'' (\texttt{i-1}) and ``bottom'' (\texttt{k-1}) neighbor. This is the basic LCD that should govern the code's runtime; the CP may be longer since it may contain instructions that are not part of the LCD. If the hardware has sufficient out-of-order capabilities, it should be able to overlap that ``extra'' part across successive loop iterations. And finally, the pure throughput prediction (TP) should be much too optimistic since it ignores all dependencies.

Since we have demonstrated OSACA's TP analysis in previous work~\cite{Osaca}, we will focus here on the refinement of runtime predictions via CP and LCD analysis. 
The total runtime is measured and combined with the number of iterations to get lattice site updates per second~[LUP/s] and cycles per iteration~[cy/it] in columns 3--4 of Table~\ref{tab:meas}.

Unrolling by the compiler must be considered when interpreting OSACA predictions since they strictly pertain to the assembly level. E.g., if a loop was unrolled four times, as it is the case for our Gauss-Seidel examples, the prediction by OSACA will be for four original (high-level) iterations. This also applies to unrolling for SIMD vectorization, which is not possible here, however.
In this paper, OSACA and IACA predictions in cycles are given for one assembly code iteration, whereas the unit ``cy/it'' always refers to high-level source code iterations.
The total unrolling factor chosen by the compilers has been 4x for all architectures.
Table~\ref{tab:gs-arm} shows the condensed OSACA output for the TX2. 
Predictions by OSACA, IACA, and LLVM-MCA can be found in Table~\ref{tab:meas}.

The predicted block throughput of all three analysis tools is far from the measurements, as expected.
Even though IACA is not capable of detecting CPs and analysing the latency of kernels anymore, its block throughput in the analysis report states 14\,cy/it, contrary to the pure port binding of 2\,cy/it. No explanation for this behavior can be found in the output although it matches exactly the LCD and the measurement.

Using the additional \texttt{-timeline} flag, LLVM-MCA provides a timeline view showing for a various number of cycles or iterations the expected cycle of dispatching, execution and retirement.
Since it models register dependencies, we assume this to be its CP analysis and expect the time from the beginning of the first iteration to the retirement of its jump instruction to be the CP length, while all further executions  have the length of the LCD.
Both numbers can be found in the last column of Tab.~\ref{tab:meas}.
While we can observe that LLVM-MCA overestimates the execution on ZEN by almost 50\%, it predicts the runtime on CLX nearly exactly.
For the ThunderX2, LLVM-MCA is neither capable of analyzing throughput nor latency at the time of writing.

OSACA provides a runtime bracket determined by the CP (upper bound) and the length of the longest cyclic LCD path (lower bound). The measured execution time should usually lie between these limits unless bottlenecks apply that are beyond our model (e.g., instruction cache misses, bank conflicts, etc.).
As seen in column 6 of Table~\ref{tab:meas}, the actual measurement lies within the prediction frame in every analysis case, and the measurement is very close to the longest LCD path for this kernel.
As expected, the runtime is faster than the pure CP length, since instructions that are not part of the LCD path  can overlap across iterations. 

\begin{figure}[tb]
\centerline{\includegraphics[width=0.85\columnwidth]{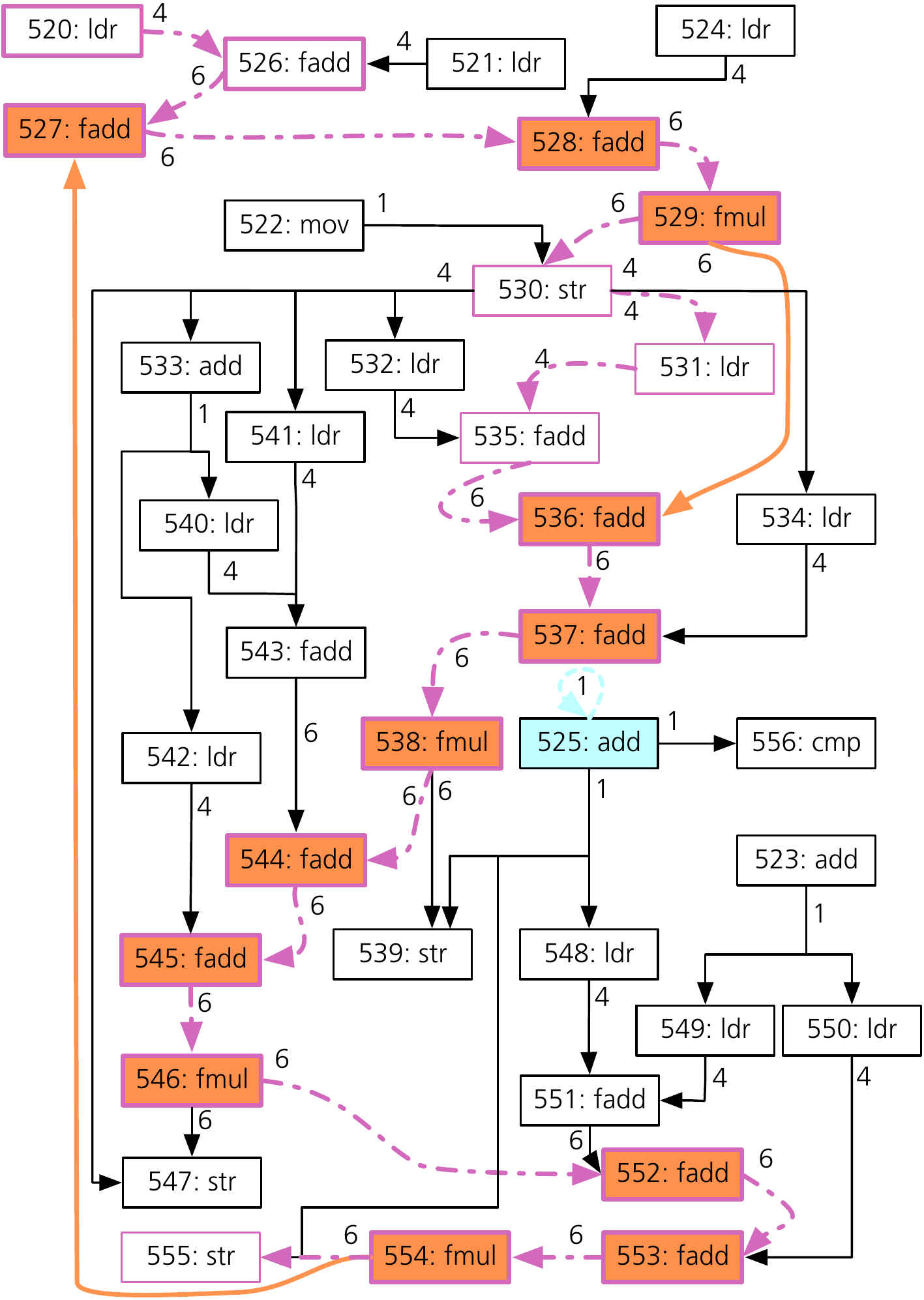}}
\caption{(Compressed) dependency graph of the Gauss-Seidel code on TX2, created by OSACA. Orange nodes are on the longest LCD, including the backedge. Pink dashed lines and outlined nodes make up the CP. Numbers in nodes are line numbers, as found in Table~\ref{tab:gs-arm}, and weights along the edges are latency cycles.}
\label{fig:dependency_graph}
\end{figure}

The detailed OSACA analysis for ThunderX2 can be found in Table~\ref{tab:gs-arm}. The LCD and CP columns show latency values for instruction forms along the CP and the longest cyclic LCD path, respectively. Fig.~\ref{fig:dependency_graph} depicts the graph generated by OSACA from the assembly.

\begin{table}[t!]
\centering
{\scriptsize
\setlength\tabcolsep{2.5pt}
\begin{tabular}{cGcGcG|cG|cl}
\hline
\rowcolor{white}P0 & P1 & P2 & P3 & P4 & P5 & LCD & CP & LN & Assembly Instructions\\
\hline
\hline
      &      &      &      &      &      &      &     & {\tiny 519} & {\tiny \texttt{.L20:}}\\
      &      &      & 0.50 & 0.50 &      &      & 4.0 & {\tiny 520} & {\tiny \texttt{ldr    d31, [x15, x18, lsl 3]}}\\
      &      &      & 0.50 & 0.50 &      &      &     & {\tiny 521} & {\tiny \texttt{ldr	d0, [x15, 8]}}\\
 0.50 & 0.50 &      &      &      &      &      &     & {\tiny 522} & {\tiny \texttt{mov	x14, x15}}\\
 0.33 & 0.33 & 0.33 &      &      &      &      &     & {\tiny 523} & {\tiny \texttt{add	x16, x15, 24}}\\
      &      &      & 0.50 & 0.50 &      &      &     & {\tiny 524} & {\tiny \texttt{ldr	d2, [x15, x30, lsl 3]}}\\
 0.33 & 0.33 & 0.33 &      &      &      &      &     & {\tiny 525} & {\tiny \texttt{add	x15, x15, 32}}\\
 0.50 & 0.50 &      &      &      &      &      & 6.0 & {\tiny 526} & {\tiny \texttt{fadd	d1, d31, d0}}\\
 0.50 & 0.50 &      &      &      &      & 6.0  & 6.0 & {\tiny 527} & {\tiny \texttt{fadd	d3, d1, d30}}\\
 0.50 & 0.50 &      &      &      &      & 6.0  & 6.0 & {\tiny 528} & {\tiny \texttt{fadd	d4, d3, d2}}\\
 0.50 & 0.50 &      &      &      &      & 6.0  & 6.0 & {\tiny 529} & {\tiny \texttt{fmul	d5, d4, d9}}\\
      &      &      & 0.50 & 0.50 & 1.00 &      & 4.0 & {\tiny 530} & {\tiny \texttt{str	d5, [x14], 8}}\\
      &      &      & 0.50 & 0.50 &      &      & 4.0 & {\tiny 531} & {\tiny \texttt{ldr	d6, [x14, x18, lsl 3]}}\\
      &      &      & 0.50 & 0.50 &      &      &     & {\tiny 532} & {\tiny \texttt{ldr	d16, [x14, 8]}}\\
 0.33 & 0.33 & 0.33 &      &      &      &      &     & {\tiny 533} & {\tiny \texttt{add	x13, x14, 8}}\\
      &      &      & 0.50 & 0.50 &      &      &     & {\tiny 534} & {\tiny \texttt{ldr	d7, [x14, x30, lsl 3]}}\\
 0.50 & 0.50 &      &      &      &      &      & 6.0 & {\tiny 535} & {\tiny \texttt{fadd	d17, d6, d16}}\\
 0.50 & 0.50 &      &      &      &      & 6.0  & 6.0 & {\tiny 536} & {\tiny \texttt{fadd	d18, d17, d5}}\\
 0.50 & 0.50 &      &      &      &      & 6.0  & 6.0 & {\tiny 537} & {\tiny \texttt{fadd	d19, d18, d7}}\\
 0.50 & 0.50 &      &      &      &      & 6.0  & 6.0 & {\tiny 538} & {\tiny \texttt{fmul	d20, d19, d9}}\\
      &      &      & 0.50 & 0.50 & 1.00 &      &     & {\tiny 539} & {\tiny \texttt{str	d20, [x15, -24]}}\\
      &      &      & 0.50 & 0.50 &      &      &     & {\tiny 540} & {\tiny \texttt{ldr	d21, [x13, x18, lsl 3]}}\\
      &      &      & 0.50 & 0.50 &      &      &     & {\tiny 541} & {\tiny \texttt{ldr	d23, [x14, 16]}}\\
      &      &      & 0.50 & 0.50 &      &      &     & {\tiny 542} & {\tiny \texttt{ldr	d22, [x13, x30, lsl 3]}}\\
 0.50 & 0.50 &      &      &      &      &      &     & {\tiny 543} & {\tiny \texttt{fadd	d24, d21, d23}}\\
 0.50 & 0.50 &      &      &      &      & 6.0  & 6.0 & {\tiny 544} & {\tiny \texttt{fadd	d25, d24, d20}}\\
 0.50 & 0.50 &      &      &      &      & 6.0  & 6.0 & {\tiny 545} & {\tiny \texttt{fadd	d26, d25, d22}}\\
 0.50 & 0.50 &      &      &      &      & 6.0  & 6.0 & {\tiny 546} & {\tiny \texttt{fmul	d27, d26, d9}}\\
      &      &      & 0.50 & 0.50 & 1.00 &      &     & {\tiny 547} & {\tiny \texttt{str	d27, [x14, 8]}}\\
      &      &      & 0.50 & 0.50 &      &      &     & {\tiny 548} & {\tiny \texttt{ldr	d30, [x15]}}\\
      &      &      & 0.50 & 0.50 &      &      &     & {\tiny 549} & {\tiny \texttt{ldr	d28, [x16, x18, lsl 3]}}\\
      &      &      & 0.50 & 0.50 &      &      &     & {\tiny 550} & {\tiny \texttt{ldr	d29, [x16, x30, lsl 3]}}\\
 0.50 & 0.50 &      &      &      &      &      &     & {\tiny 551} & {\tiny \texttt{fadd	d31, d28, d30}}\\
 0.50 & 0.50 &      &      &      &      & 6.0  & 6.0 & {\tiny 552} & {\tiny \texttt{fadd	d2, d31, d27}}\\
 0.50 & 0.50 &      &      &      &      & 6.0  & 6.0 & {\tiny 553} & {\tiny \texttt{fadd	d0, d2, d29}}\\
 0.50 & 0.50 &      &      &      &      & 6.0  & 6.0 & {\tiny 554} & {\tiny \texttt{fmul	d30, d0, d9}}\\
      &      &      & 0.50 & 0.50 & 1.00 &      & 4.0 & {\tiny 555} & {\tiny \texttt{str	d30, [x15, -8]}}\\
 0.33 & 0.33 &      &      &      &      &      &     & {\tiny 556} & {\tiny \texttt{cmp	x7, x15}}\\
      &      &      &      &      &      &      &     & {\tiny 557} & {\tiny \texttt{bne	.L20}}\\
\hline
\rowcolor{white} 9.83 & 9.83 & 1.33 & 8.00 & 8.00 & 4.00 & 72.0 & 100.0 & \multicolumn{2}{l}{sum (4x unrolled)} \\
\rowcolor{white} \textbf{2.46} & 2.46 & 0.33 & 2.00 & 2.00 & 1.00 & \textbf{18.0} & \textbf{25.0} & \multicolumn{2}{l}{per high-level iteration} \\
\hline
\end{tabular}}
\caption{(Condensed) OSACA analysis of Gauss-Seidel assembly code for ARM-based ThunderX2 architecture. The LN column are line numbers.}
\label{tab:gs-arm}
\end{table}

Note that in cases where the LCD is very short or zero, the throughput prediction applies, and
a deviation of the measurement from this lower limit points to either a shortage of OoO resources (physical registers, reorder buffer) or an architectural effect not covered by the machine model.

\subsection{Validation Hardware, Software, and Runtime Environment}\label{ssec:validation_hardware}

OSACA (version 0.3.1.dev0) was run with Python~v3.6.8 and benchmarks were compiled using Intel ifort~v19.0.2 and GNU Fortran (ARM-build-8) 8.2.0, respectively. 
All results presented were gathered on three machines, with fixed clock frequency and disabled turbo mode:
\paragraph*{ThunderX2}
ARM-based Marvell ThunderX2 9980 with ThunderX2 micro-architecture (formerly known as Cavium Vulcan) at 2.2\,GHz (TX2), gfortran, options \texttt{-mcpu=thunderx2t99+simd+fp -fopenmp-simd -funroll-loops -Ofast}
\paragraph*{Cascade Lake} Intel Xeon Gold 6248 with Cascade Lake X micro-architecture at 2.5\,GHz (CLX), ifort, options \texttt{-funroll-loops -xCASCADELAKE -Ofast}
\paragraph*{Zen} AMD EPYC 7451 with Zen micro-architecture at 2.3\,GHz (ZEN), gfortran, options \texttt{-funroll-loops -mavx2 -mfma -Ofast}

The process was always bound to a physical core.
In effect, statistical runtime variations were small enough to be ignored.

\section{Conclusion}\label{sec:conclusion}

\subsection{Summary}
We have shown that automatic extraction, throughput, and critical path analysis of assembly loop kernels is feasible using our cross-platform tool OSACA.
OSACA’s results are accurate and sometimes even more precise and versatile than predictions of comparable tools like IACA and LLVM-MCA.
Additionally, direct critical path analysis including loop-carried dependencies is not supported by any other tool to date, although it can be inferred manually from LLVM-MCA's timeline information.

\subsection{Future Work}\label{ssec:future}
In the future we intend to extend OSACA to support hidden dependencies, i.e., instructions accessing resources not named specifically in the assembly, such as status flags and load-after-store dependencies, including stack operations.
The LCD analysis is not perfect and may miss dependencies in some special cases, which can be improved by taking more than two iterations into account.
Furthermore, we plan to increase the number of micro-benchmark interfaces and to support the semi-automatic usage of asmbench in the OSACA toolchain.
Beyond the even distribution of \muops\ across multiple ports, we want to implement a more realistic scheduling scheme that takes port utilization into account. Support for new micro-architectures like AMD's Zen~2 and eventually IBM's Power9 is also planned.
Another topic is the overlap of latency in complex instructions, which can change the outcome of the analysis slightly but may be significant in pathological cases (e.g., in $a = a + b\times c$ with an FMA instruction, the multiplication may already execute before $a$ becomes available). The split, as well as the fusion, of \muops\ is currently not considered, but can be achieved with replacement rules in the architecture model description.

Accurately modeling the performance characteristics of the decode, reorder buffer, register allocation/renaming, retirement and other stages, which all may limit the execution throughput and impose latency penalties, is currently out of scope for OSACA.

\section*{} 

\bibliographystyle{IEEEtran}
\bibliography{IEEEabrv,citations}

\end{document}